\documentclass[12pt,a4paper]{article}
\usepackage{graphicx}
\usepackage{geometry}
\begin{document}
\title{Monte Carlo aided design of the inner muon veto detectors for the Double Chooz experiment}

\author{D. Dietrich\thanks{Corresponding author.}~, 
D. Greiner, 
J. Jochum, \\
T. Lachenmaier,
M. R\"{o}hling, 
L. F. F. Stokes\\
\llap Kepler Center for Astro and Particle Physics,\\
Auf der Morgenstelle 14, 72076 T\"{u}bingen, Germany\\
E-mail: dennis.dietrich@uni-tuebingen.de}
\maketitle

\abstract{The Double Chooz neutrino experiment aims to measure the last unknown neutrino mixing angle $\theta_{13}$ using 
two identical detectors positioned at sites both near and far from the reactor cores of the Chooz nuclear power plant. 
To suppress correlated background induced by cosmic muons in the detectors, they are protected by veto detector systems. 
One of these systems is the inner muon veto. It is an active liquid scintillator based detector and instrumented with 
encapsulated photomultiplier tubes. In this paper we describe the Monte Carlo aided design process of the inner muon veto, that 
resulted in a detector configuration with 78 PMTs yielding an efficiency of $99.978\pm 0.004$\,\% for 
rejecting muon events and an efficiency of $>98.98$\% for rejecting correlated events induced by muons. A veto detector of this 
design is currently used at the far detector site and will be built and incorporated as the muon identification system at the 
near site of the Double Chooz experiment.} 

\section{Introduction}
\label{section_introduction}

The Double Chooz experiment \cite{DCProposal} is a neutrino oscillation experiment located at the Chooz nuclear power plant 
in France. The experiment will use two identical neutrino detectors, a near one (400\,m away from the reactor cores) and 
a far one (1.05\,km away) for a relative measurement of the neutrino fluxes of the reactor cores. The far detector is currently 
taking data. The near detector lab is currently being excavated. Data taking with both detectors is planned to start 
in 2013. The experiment aims to accurately measure the third neutrino mixing angle $\theta_{13}$ of the PMNS-matrix \cite{MNS,Ponte}. 
A first indication of a non-zero mixing angle $\theta_{13}$ of $\sin^2(2\theta_{13})=0.086\pm0.041(stat.)\pm0.030(syst.)$ was obtained 
with the far detector only \cite{DC1stPub}.\\

\begin{figure}[t]
        \begin{center}
                \includegraphics[width=15cm]{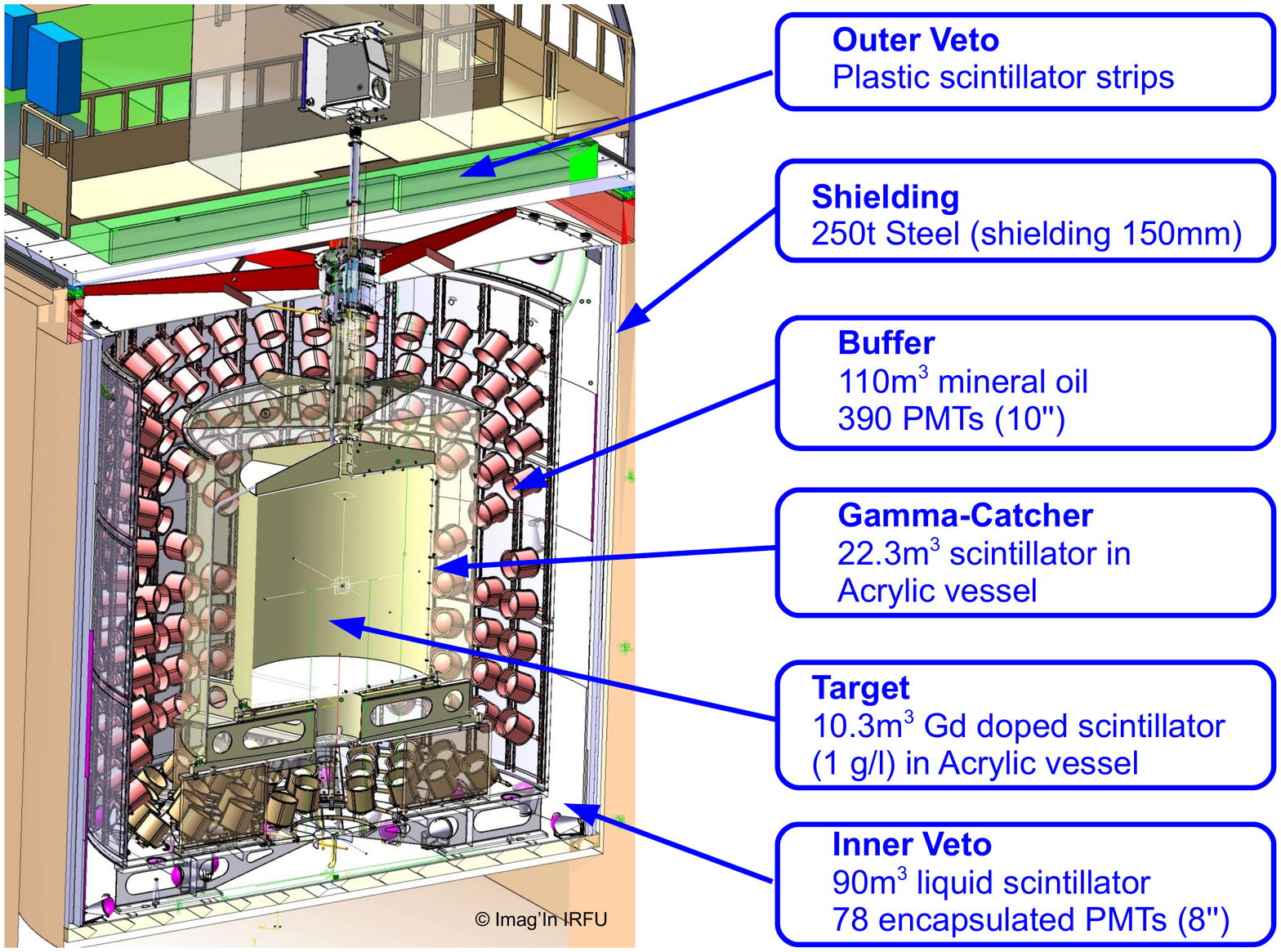}
                \caption{Schematic overview of the Double Chooz multi detector concept. Image courtesy of CEA IRFU, France.
A cross section of the detector and its dimensions can be found in \cite{DC1stPub}.}
                \label{geometry}
        \end{center}
\end{figure}

Both detectors consist of concentric cylindrical volumes filled with liquids. A schematic overview of the Double Chooz detector design 
\cite{DCProposal,Greiner:2007zzd,DCDetector} is shown in figure \ref{geometry}. The innermost volume is the neutrino target (NT) which 
is filled with a Gadolinium doped ($1$\,g/l Gd) liquid scintillator \cite{Buck}. Electron anti-neutrinos are detected via the inverse 
$\beta$-decay reaction $\bar\nu_e+p\rightarrow e^++n$ and the subsequent neutron capture on Gd in the NT. This gives a clear experimental 
signature, first the ``prompt'' positron energy deposition and after a mean time of 30\,$\mu$s the ``delayed'' neutron capture with an 
8\,MeV $\gamma$-ray cascade. The gamma-catcher (GC) encloses the NT and is filled with an undoped liquid scintillator. Its purpose is to 
detect $\gamma$-rays from the neutron capture escaping the target region. A buffer volume filled with non-scintillating liquid surrounds 
these active volumes. $390$ $10$-inch photomultiplier tubes (details can be found in \cite{BufferPMTs1,BufferPMTs2}) are mounted inside 
the buffer to collect the scintillation light from the NT and GC. The buffer volume acts mainly to suppress the radioactivity from the 
photomultiplier (PMT) glass. These innermost volumes, the NT, GC and buffer form the inner detector (ID).\\

The ID is surrounded by the inner muon veto (IV), a liquid scintillator detector instrumented with 
encapsulated 8-inch PMTs. The IV has a height of 7\,m and the distance between the buffer vessel and the veto vessel is $50$\,cm. 
This sub-detector is designed to allow the tagging of muons passing through or near the active neutrino detector volume with very high efficiency. 
These muons can create spallation neutrons, which may mimic the prompt and delayed part of a neutrino signal via proton recoils and subsequent 
capture on gadolinium and are therefore a potentially dangerous background. The veto tank itself is shielded by iron bars in the case of the 
far detector and by water in case of the near detector. This shielding aims to prevent gammas from rock radioactivity entering the inner parts 
of the detector. \\

The reduction and proper understanding of the background is a key factor in order to achieve the required
sensitivity for measuring $\theta_{13}$. The Double Chooz detectors will be operated at shallow depths and even with this shielding 
the rate of cosmic muons that enter the detectors is still high (see table \ref{table_NearFarSite}). These cosmic muons produce unstable 
isotopes ($^9$Li and $^8$He) as well as protons or neutrons via spallation or due to capture reactions. This can mimic the prompt positron 
signal before a neutron is captured on Gd.  Hence the main purpose of the IV lies within its ability to tag muons crossing the detector 
volumes and to reduce correlated background. Complementary to the IV there is an Outer muon Veto (OV), which consists of two layers of 
plastic scintillator strips covering the top of the detector system (different areas for near and far sites). The Monte Carlo studies in 
this paper do not use the OV. But this system is expected to provide additional suppression of events induced by cosmic muons.

\begin{table}[htbp]
  \centering
    \begin{tabular}{lcc}\hline\hline
                          &near site    &far site\\
      depth (m.w.e.)      &115          &300\\
      $E_\mu$ (GeV)       &30           &61\\
      $R_\mu^{IV}$ (Hz)   &440          &40 \\\hline\hline
   \end{tabular}
  \caption[Comparison near versus far detector site]{Comparison of depth, average muon energy $E_\mu$ and expected
muon rate in the IV $R_\mu^{IV}$ at the near site versus the far site. The values for the muon rate and mean muon energy at the
far site are extracted from \cite{MuonSim} and from Monte Carlo simulations described in the proposal \cite{DCProposal} at
the near site with updated values of the detector position respectively.}
  \label{table_NearFarSite}
\end{table}

\section{Simulation input parameters}
\label{section_requirements}

In the first part of this work different PMT layouts and numbers were evaluated. Within both IVs each PMT will 
be encapsulated in a stainless steel cone. It widens conically and is closed to the outside by a transparent window (design 
similar to \cite{BorexinoEncap2}). The PMTs are 8-inch Hamamatsu R1408 \cite{R1408Manual}, with a combined 
quantum and collection efficiency of 20\% in the relevant wavelength region. The scintillating liquid was developed by the 
Technische Universit\"{a}t M\"{u}nchen, Germany. It consists of the aromatic compound Linear Alkyl Benzene (LAB) and 
n-tetradecane. In our simulation the light yield is 8500\, photoelectrons (PE) per MeV for electrons. 
There is no experimental data on quenching factors for this scintillator composition. The quenching factors 
implemented for electrons:alpha-particles:protons are 1:5:10. The design process of the IV was extensively aided by 
the use of Monte Carlo simulations to ensure feasibility and in order to optimize the detector layout. The dimensions of the 
veto vessel, the specifications of the PMTs as well as the characteristics of the IV liquid scintillator and the muon spectra 
and muon angular distribution were given as an input into the Monte Carlo simulation.\\
 
The Monte Carlo simulation used within the Double Chooz collaboration is based on the GEANT4 (version 4.9.2) 
simulation framework \cite{Geant4} and the generic liquid scintillator neutrino experiment package GLG4sim \cite{GLG4sim}. 
The second part of this work is devoted to the determination of the efficiencies for rejecting muons and correlated 
events induced by them. Here the QGSP\_BIN\_HP Physics List was used for the neutron production by muons. In addition, for muon 
propagation through rock the tool MUSIC \cite{MUSIC} was utilised to obtain muon momenta at the 
detector underground sites.\\

\section{Design studies using Monte Carlo simulations}
\label{section_design}

\begin{figure}[]
\begin{center}
$\begin{array}{cccc}
        \mathrm{60\,PMTs} & \mathrm{72\,PMTs} & \mathrm{78\,PMTs} & \mathrm{108\,PMTs}\\ 
        \includegraphics[width=3.0cm]{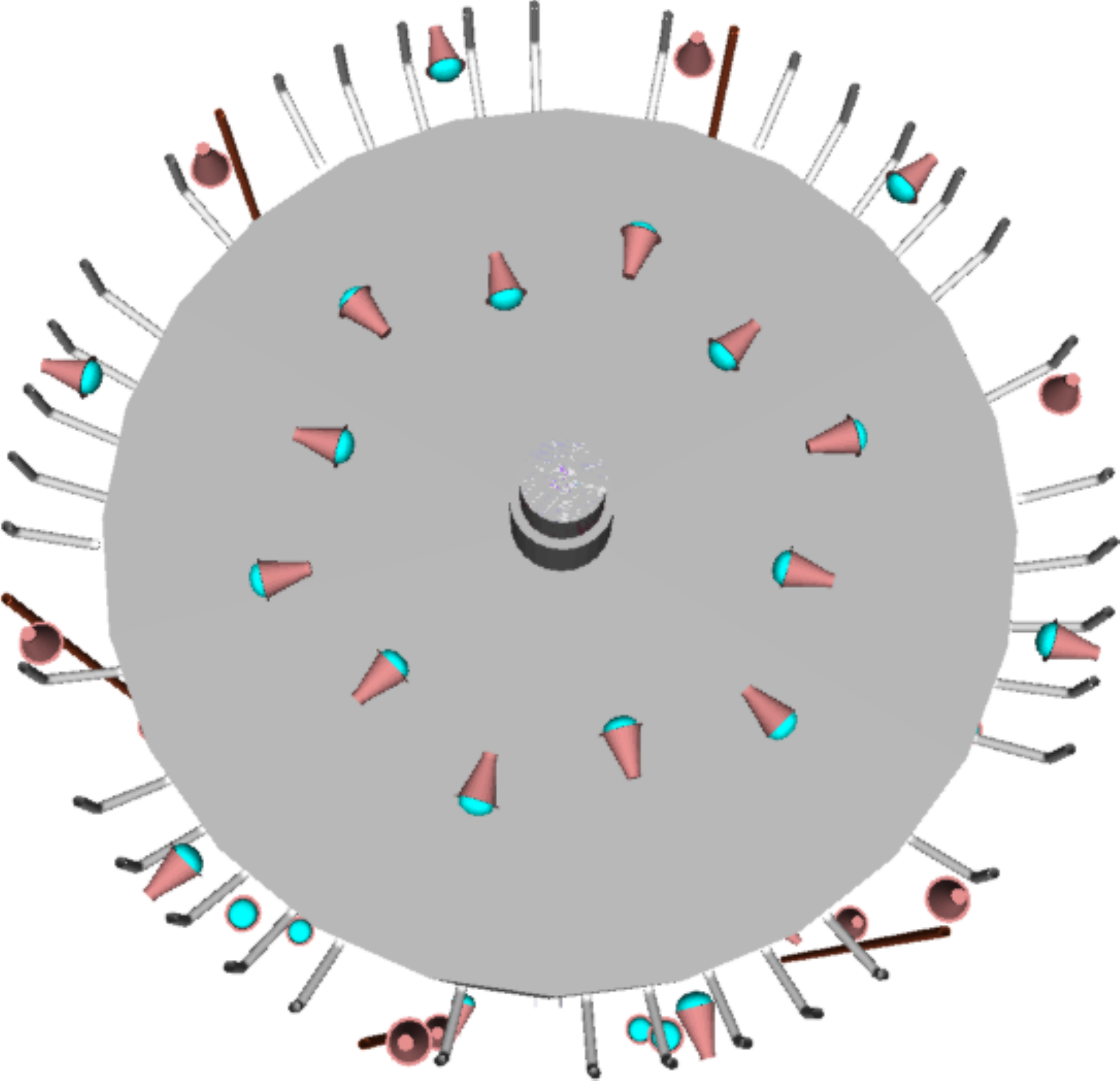}&
        \includegraphics[width=3.0cm]{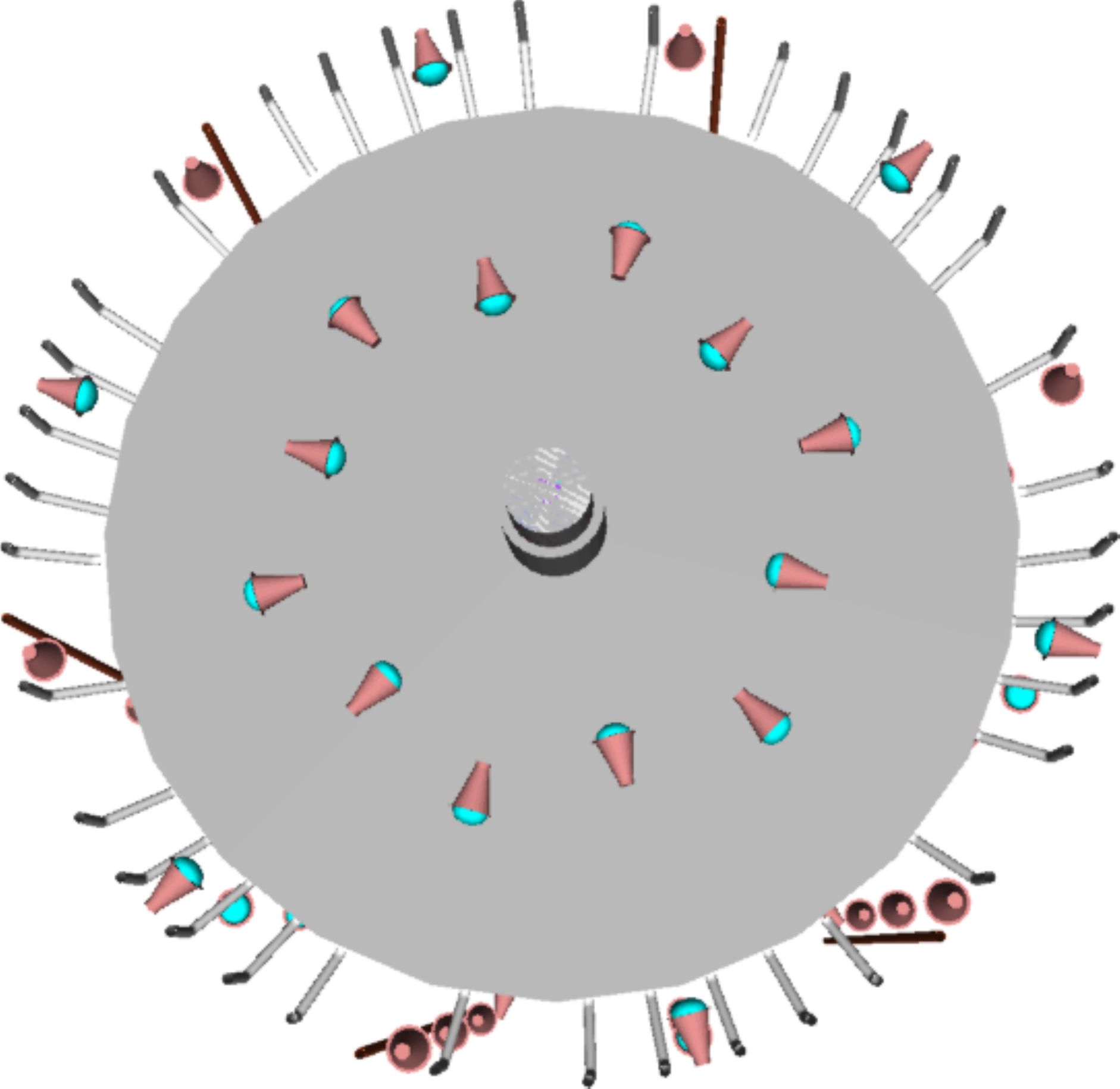}&
        \includegraphics[width=3.0cm]{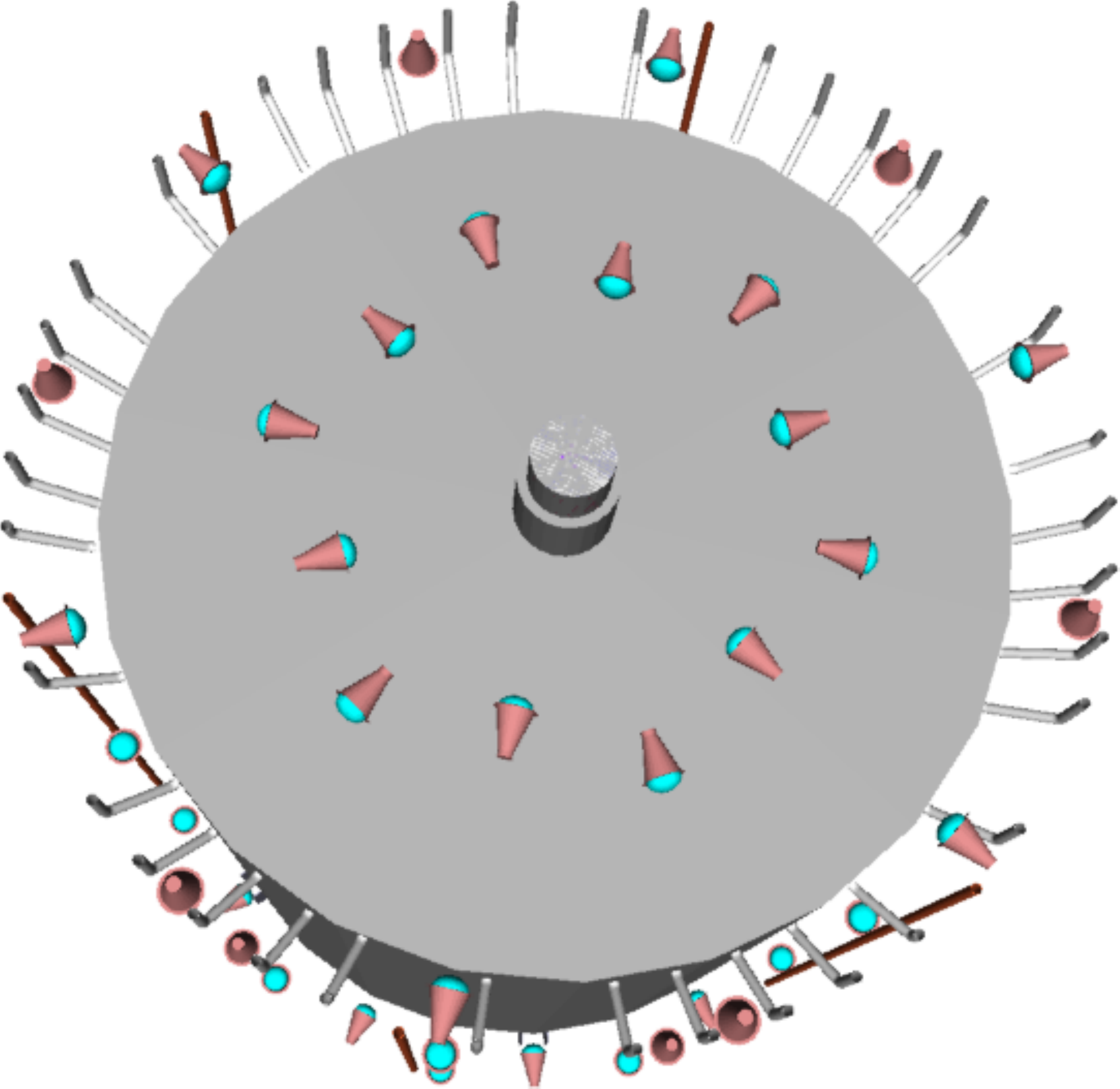}&
        \includegraphics[width=3.0cm]{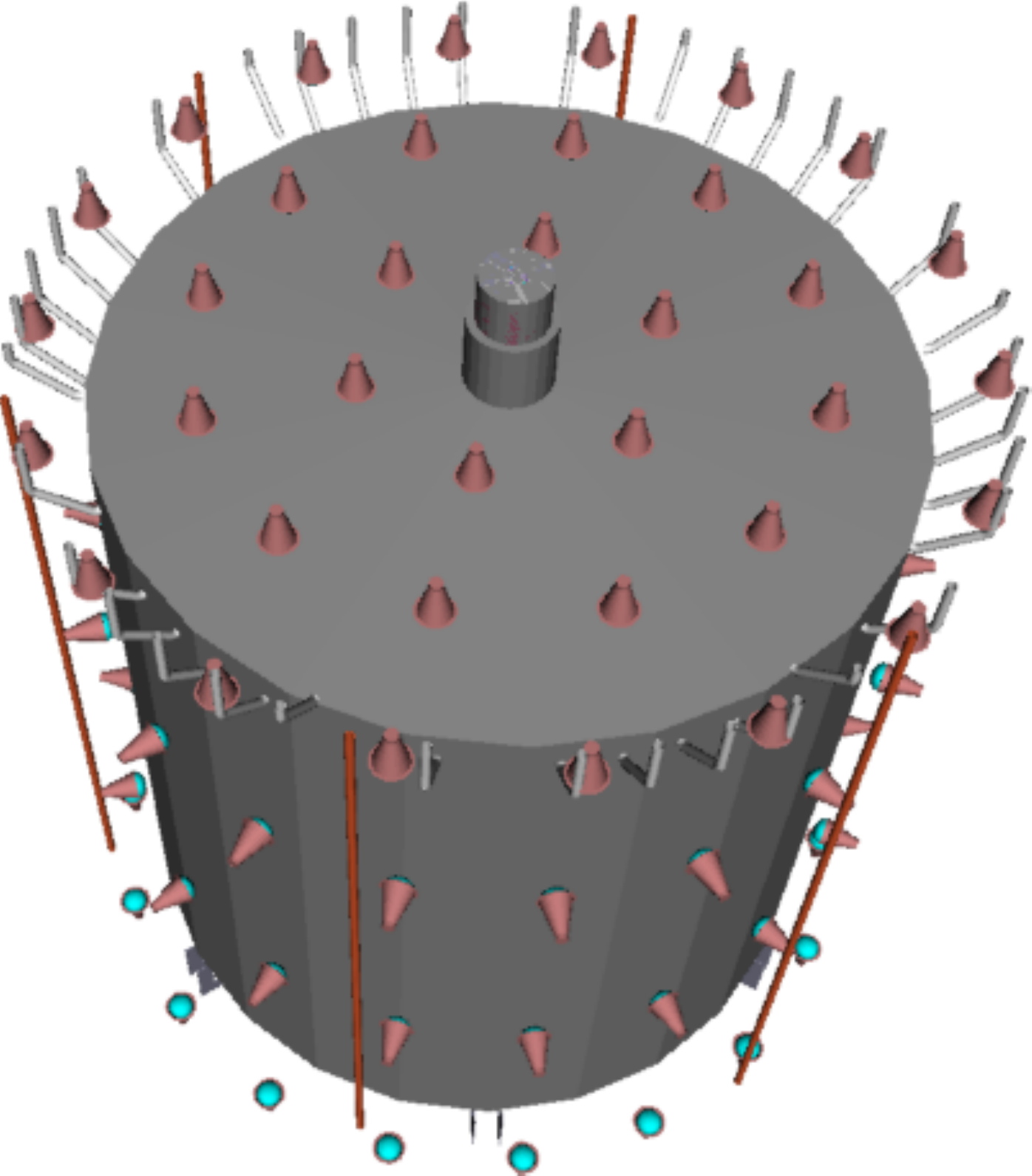}\\          
                        
        \includegraphics[width=3.0cm]{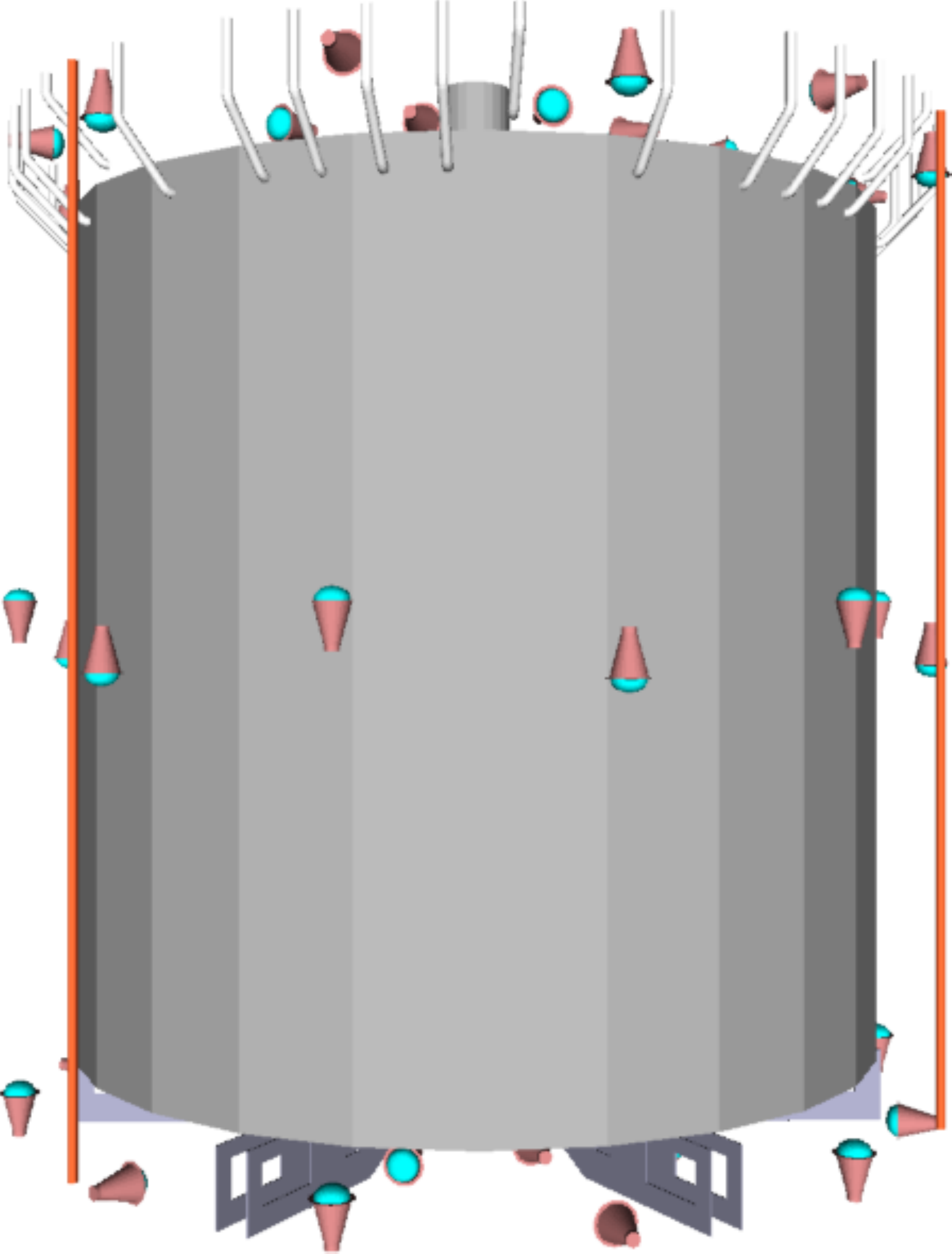}&
        \includegraphics[width=3.0cm]{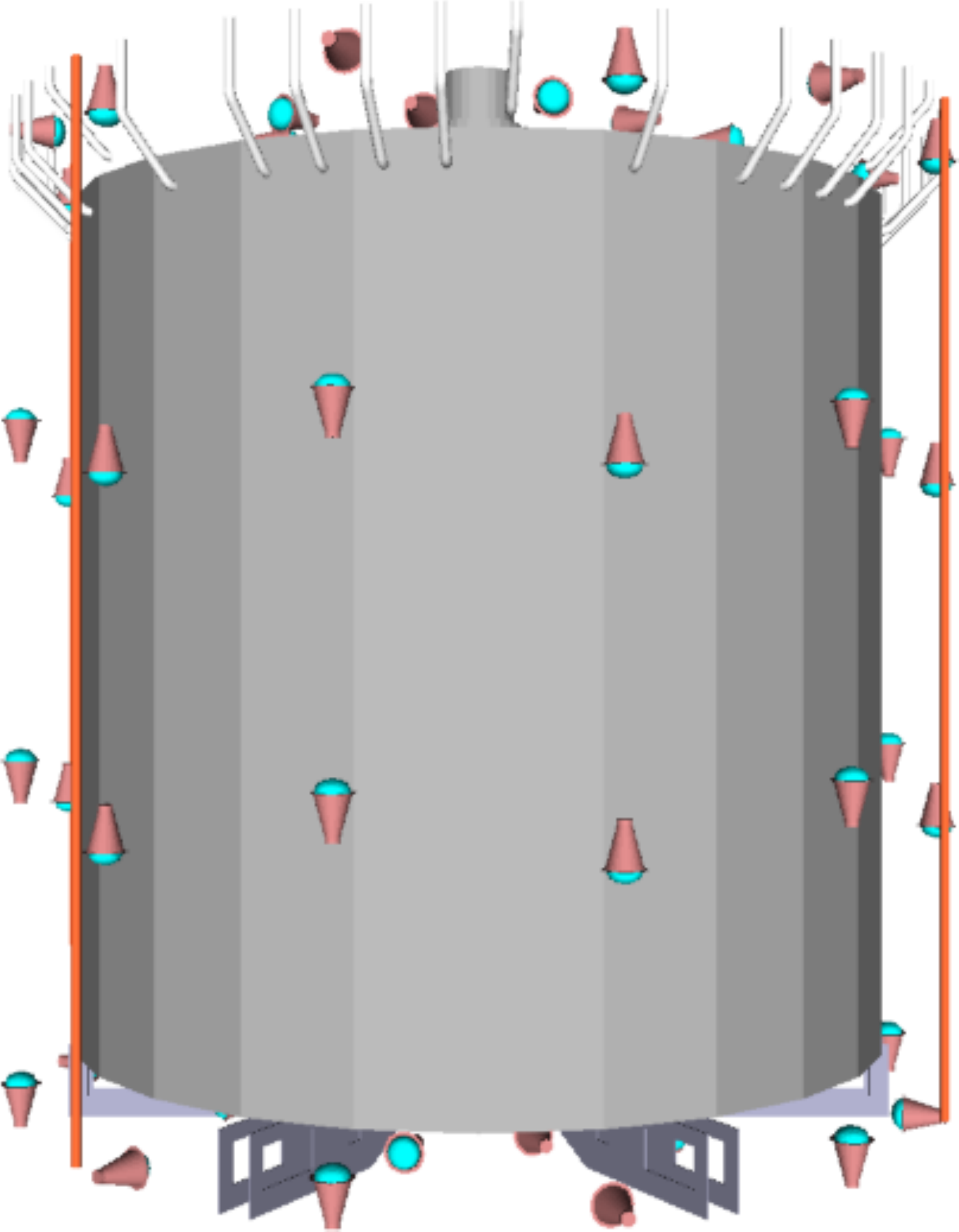}&        
        \includegraphics[width=3.0cm]{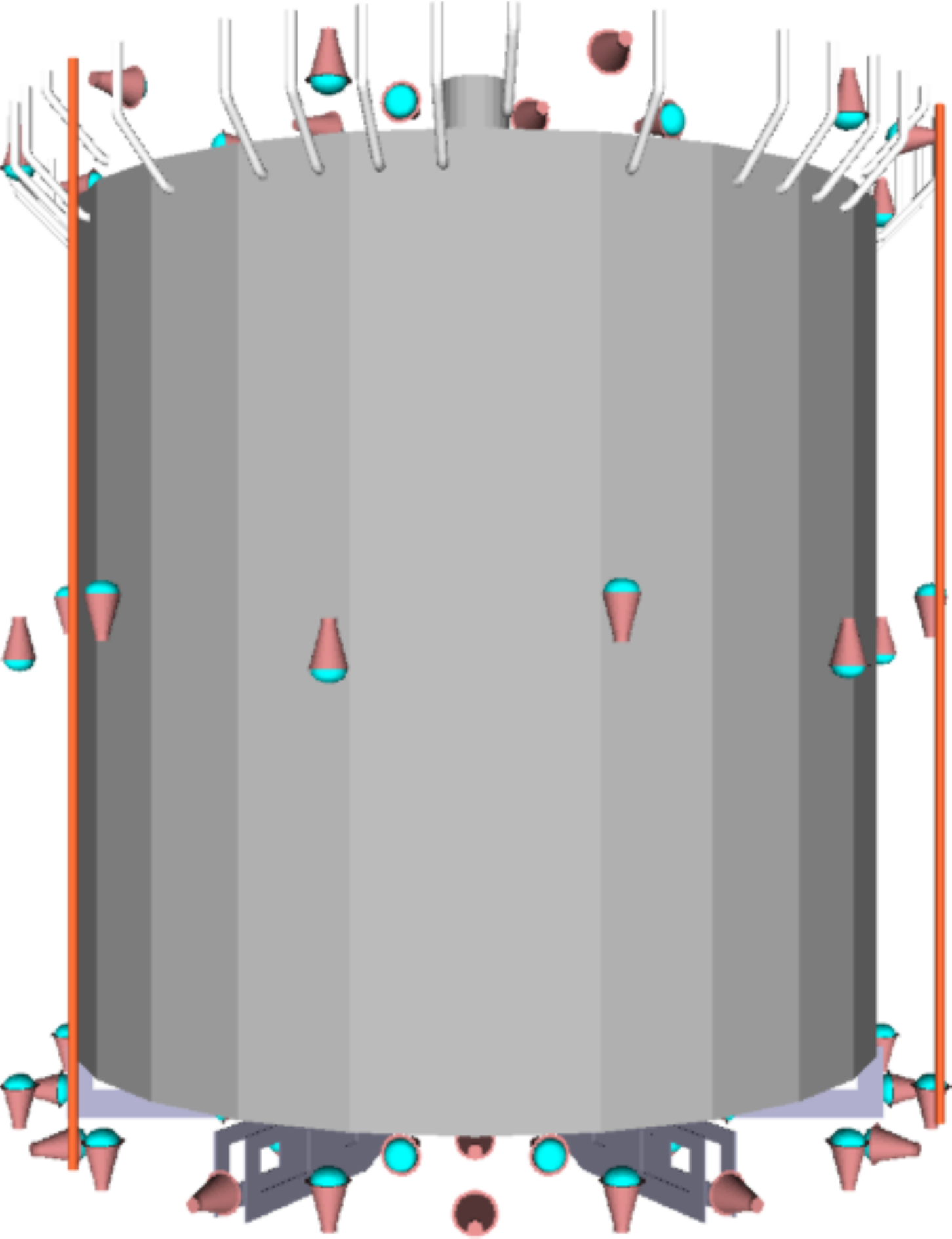}&
        \includegraphics[width=3.0cm]{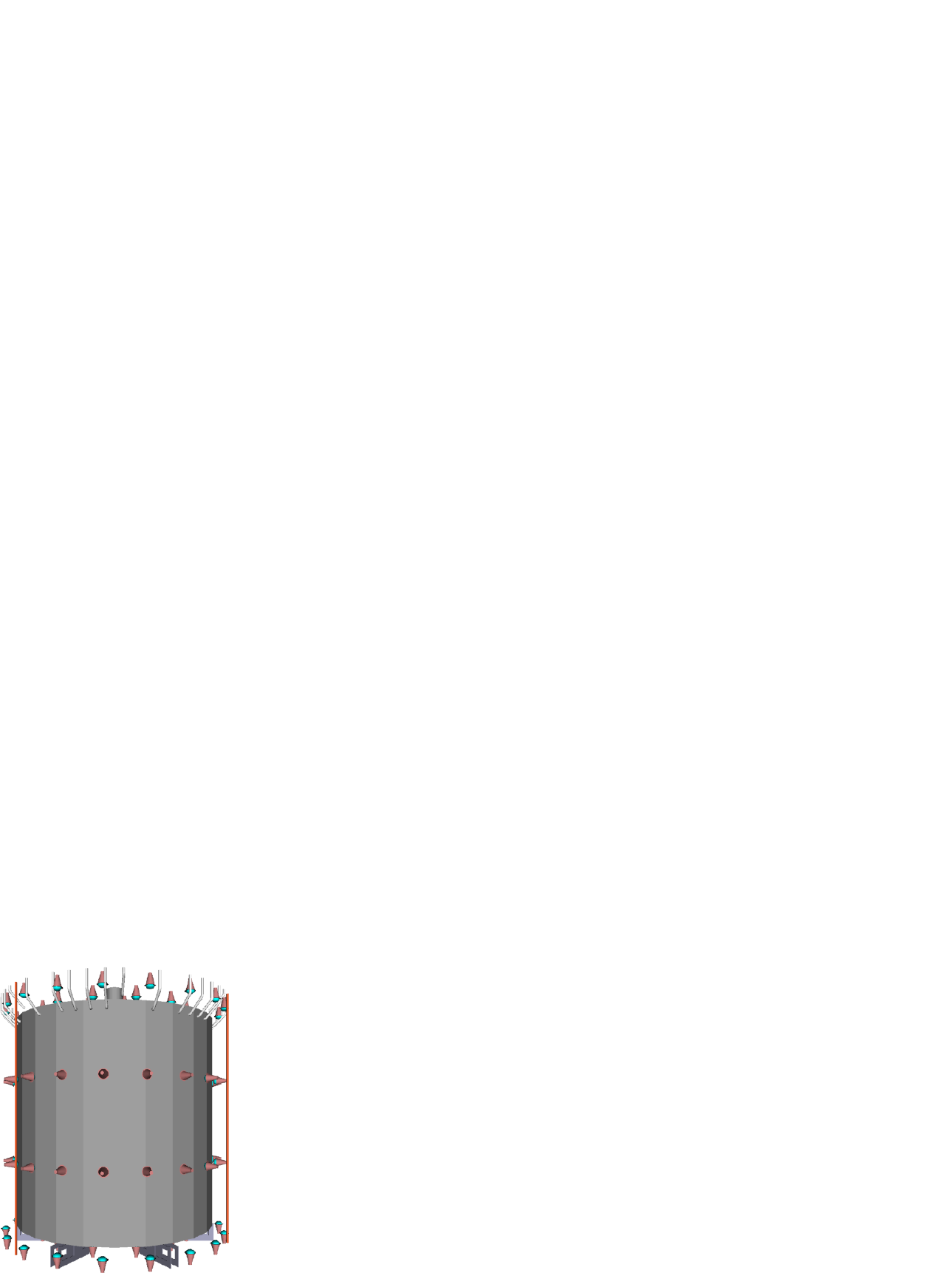}\\        
                
        \includegraphics[width=3.0cm]{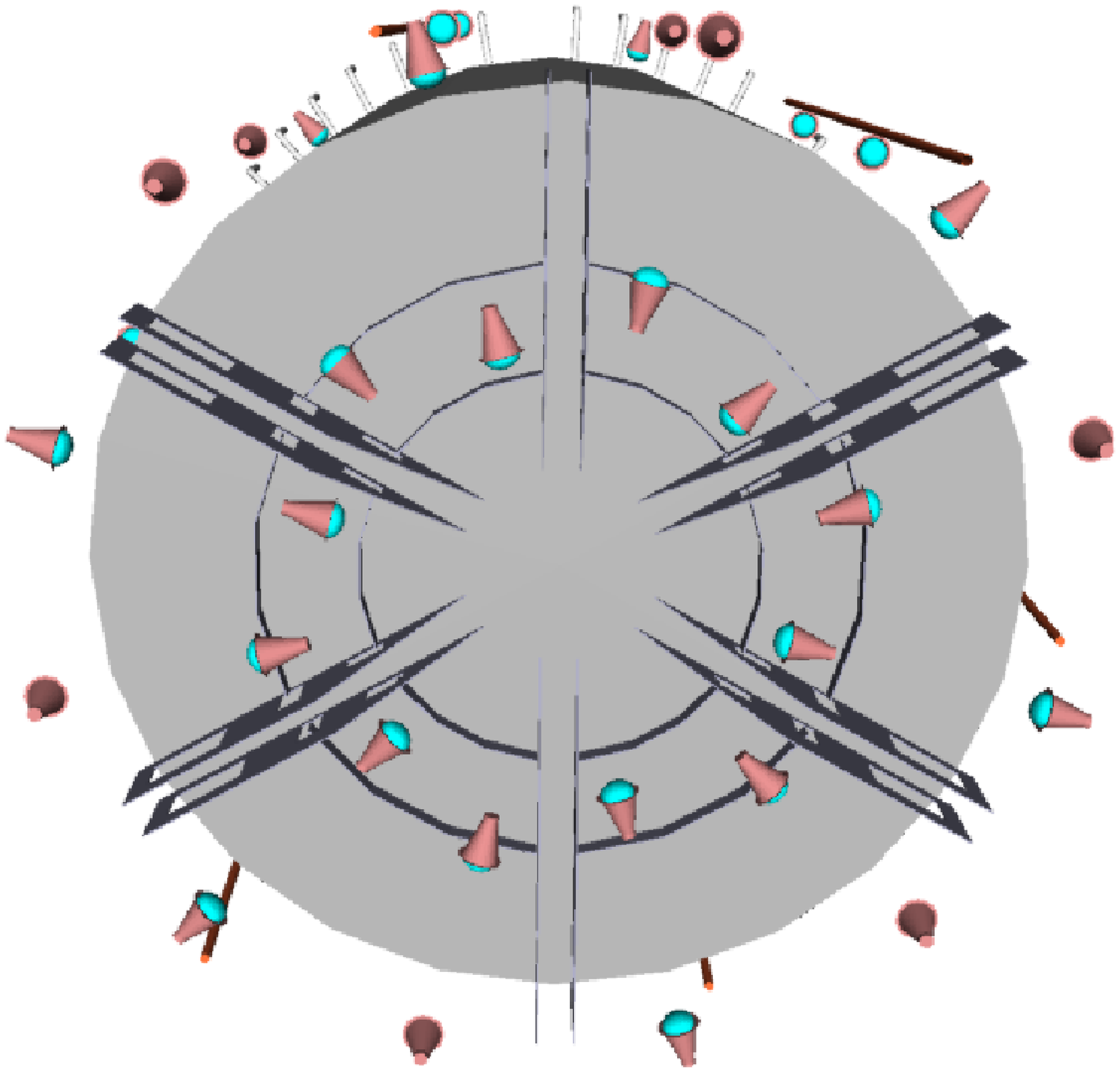}&
        \includegraphics[width=3.0cm]{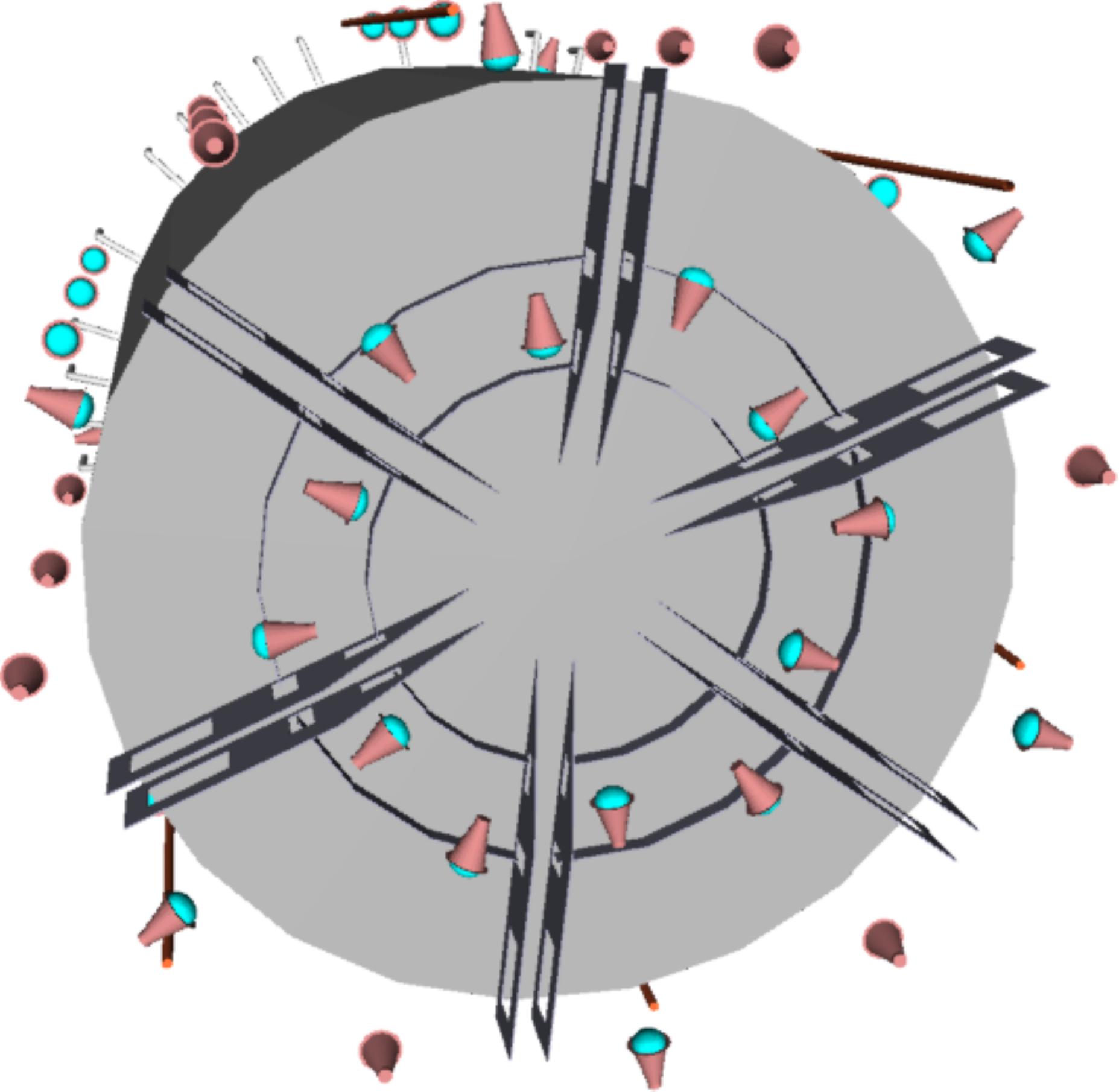}&
        \includegraphics[width=3.0cm]{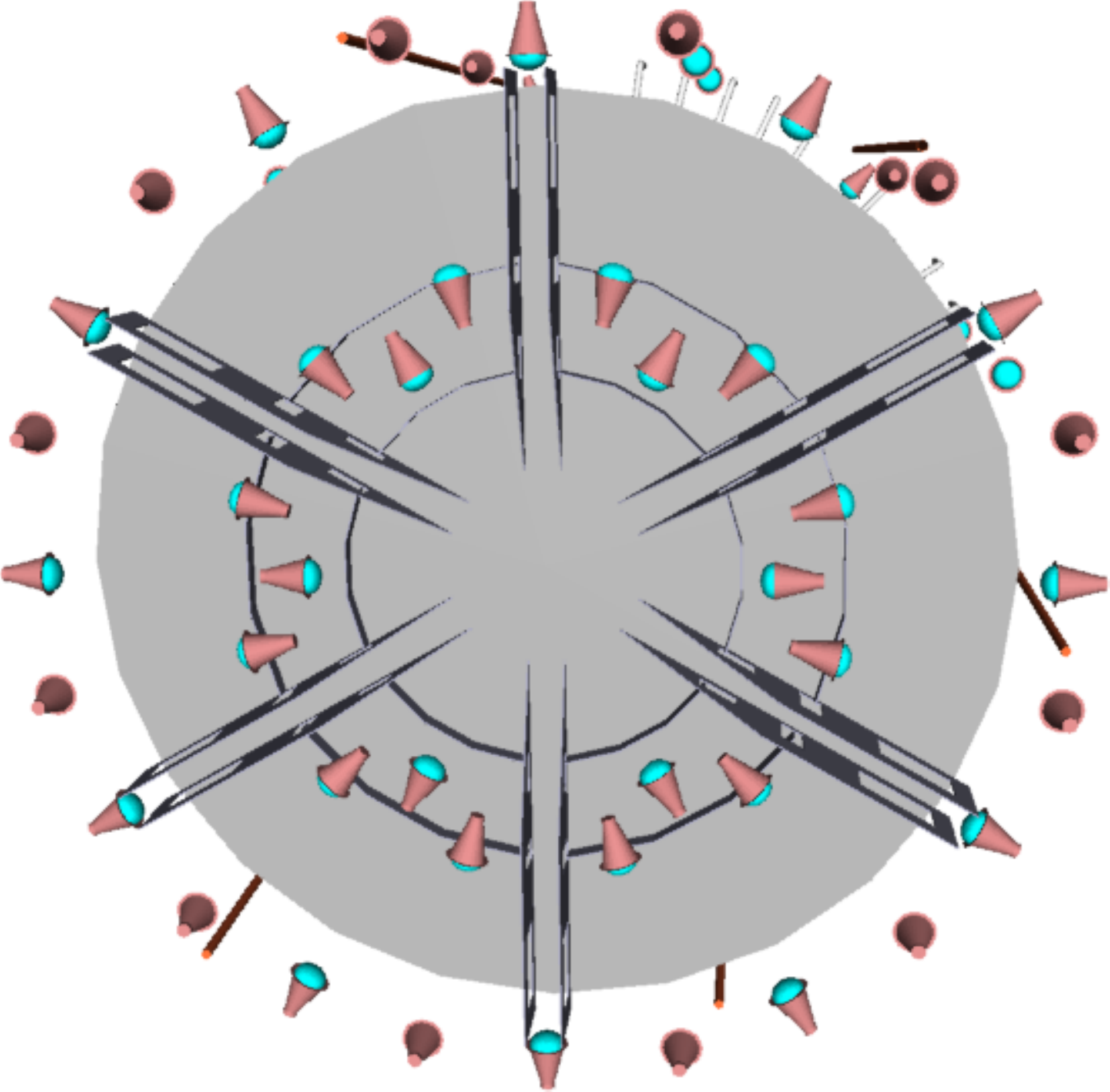}&
        \includegraphics[width=3.0cm]{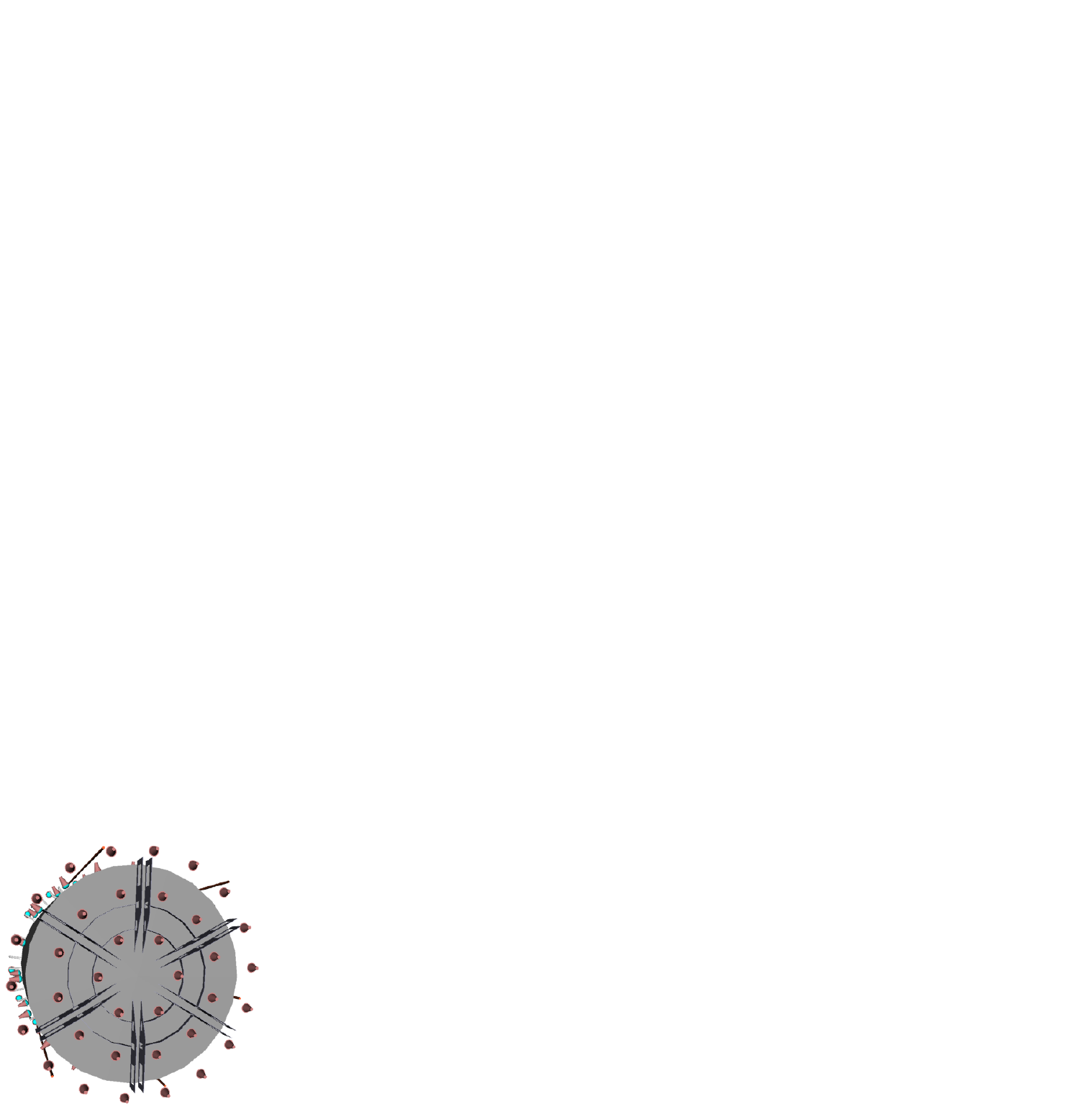}
\end{array}$
\end{center}
\caption{PMT distributions implemented in the simulations showing the top, side and bottom for the configurations with 
60, 72, 78 and 108 PMTs. The interior grey vessel is the buffer, the veto vessel itself is not shown. 
The MC implementation of the IV's geometry includes various pipes and support structures that are within the IV vessel 
and can create shadowing effect for the PMTs.}
\label{fig:Layouts}
\end{figure}

The starting point of the simulations was a sensible distribution of the PMT positions with 60 PMTs.
Local energy depositions of electrons with a kinetic energy of 1\,MeV were uniformly distributed throughout the IV volume.
The mean number of photoelectrons per MeV was extracted and after this, PMTs were added or moved in multiples of six
(corresponding to the 6-fold geometry induced by the support structure of the buffer vessel) and the simulation was repeated
with the new PMT positions. In iterating this, the PMT distribution was optimized to have the most homogeneous response and a 
sufficient amount of light collected in terms of photoelectrons/MeV (PE/MeV). This number should be as high as possible so that muons which only clip the corners of the detector and thus have a short track-length in the IV produce enough photoelectrons 
to be detected. Figure \ref{fig:Layouts} shows the layouts considered with 60, 72, 78 and 108 PMTs within the GEANT4 simulation. 
The configuration with 108 PMTs is special, because all PMTs look towards the buffer vessel. Whereas the PMTs in all the other 
configurations alternate facing upwards and downwards on the sides, or inwards and outwards on top of or beneath the buffer vessel.\\

\begin{figure}
\centering
\begin{minipage}[t]{0.49\textwidth}
\includegraphics[width=\textwidth]{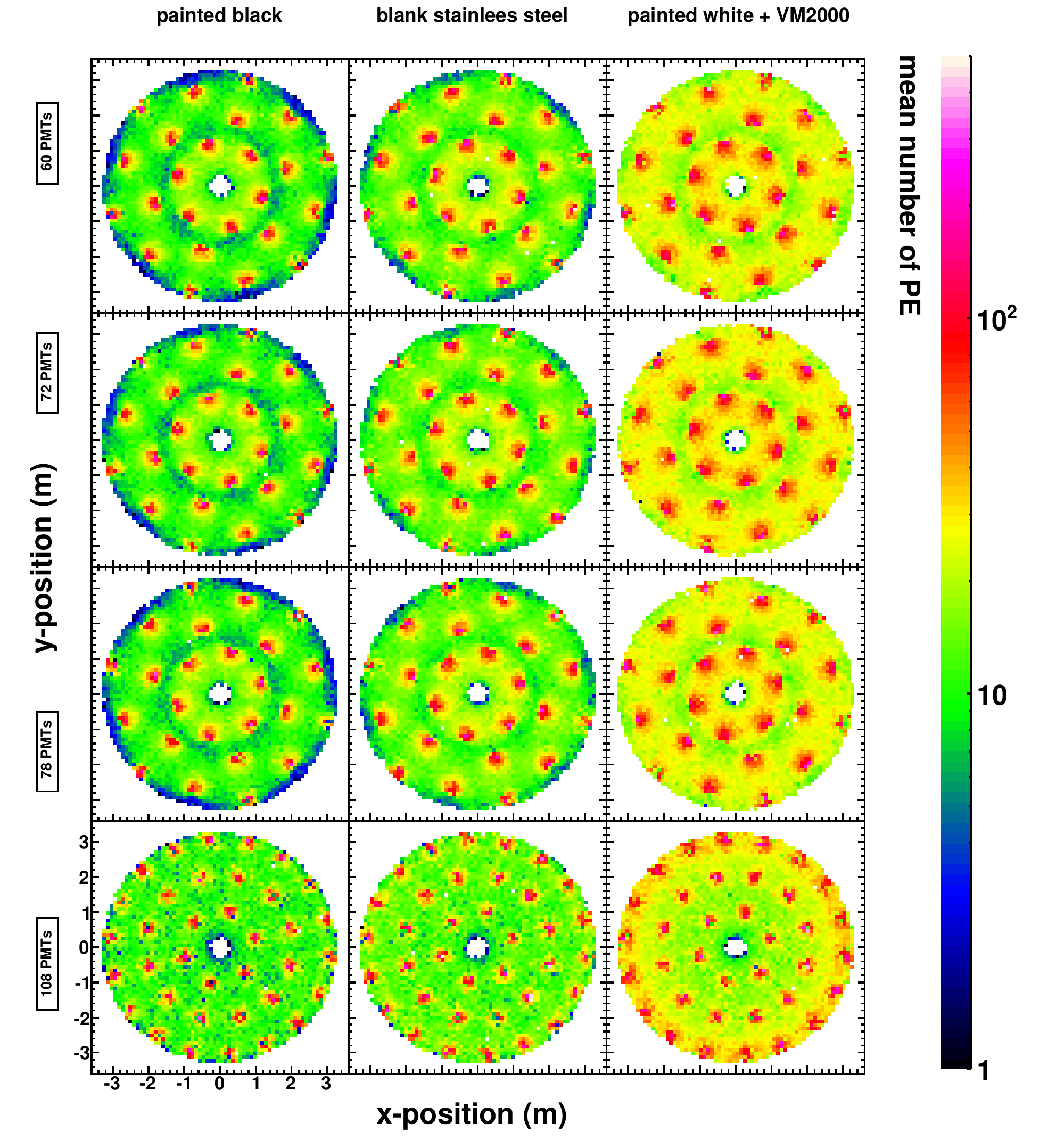}
\end{minipage}
\begin{minipage}[t]{0.49\textwidth}
\includegraphics[width=\textwidth]{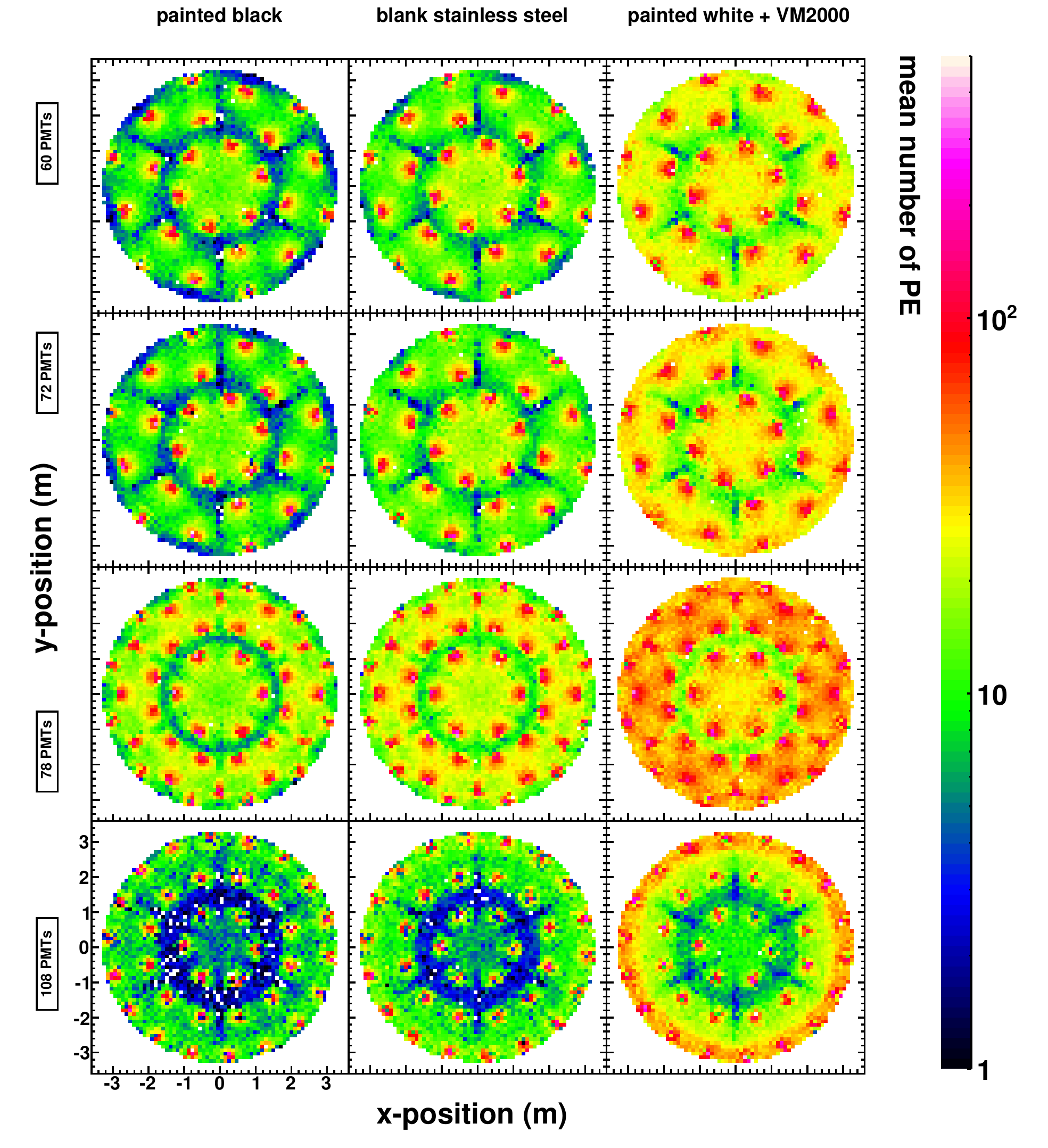}
\end{minipage}
\begin{minipage}[t]{0.5\textwidth}
\includegraphics[width=\textwidth]{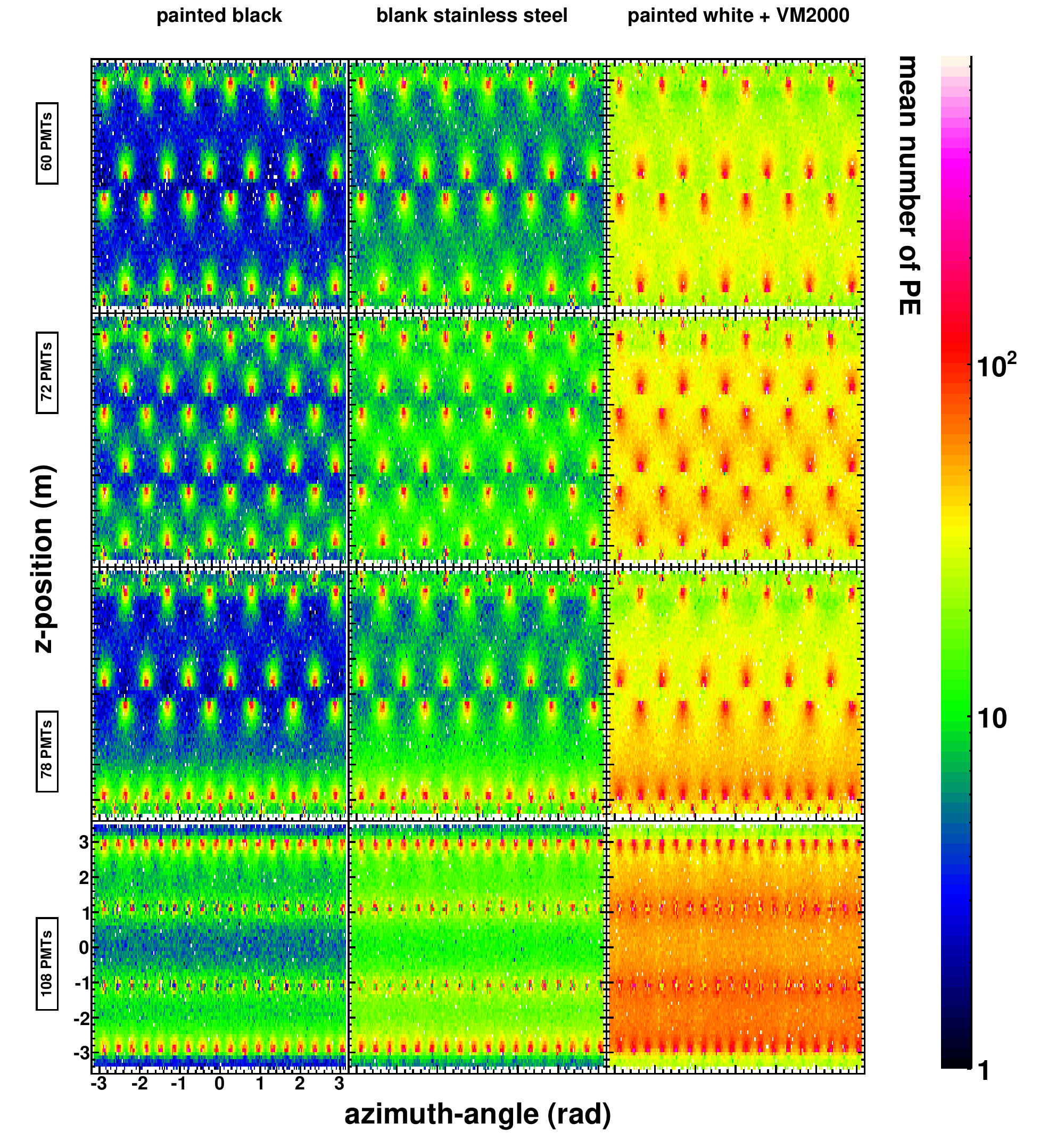}
\end{minipage}
\caption{Map of PE/MeV summed over all PMTs for 1\,MeV electrons in the top (left plot), bottom (right plot) and 
side (bottom plot) of the IV. The option with white painted walls and VM2000 foil has the highest number of PE/MeV. This option with 
78 PMTs was chosen as it shows a high value of PE/MeV and minimising shadowing effects.}
\label{fig:MeanPEPlots}
\end{figure}

The influencing factors for the light collection efficiency are the geometrical optical coverage, defined by the IV geometry, the 
absorption probability of the scintillating liquid, reflections on the IV vessel surface, as well as the quantum efficiency of the PMTs. 
For photons with a wavelength of $420$\,nm, which corresponds to the maximum of the scintillator emission spectrum, the attenuation 
length was assumed to be $9.4$\,m in the IV scintillating liquid. As the width ($\approx 50$\,cm) of the IV is small compared to the 
attenuation length the reflectivity of the surfaces has a strong influence on the detection efficiency. Three options were considered 
for the IV vessels' surfaces: first blank stainless steel walls, second black painted walls and thirdly white painted walls 
and highly reflective foil (VM2000 from 3M) attached to the buffer vessel walls. In the last two options only the IV vessel is painted, 
whereas the buffer vessel surface is left as it is (stainless steel) for the black wall option and coated with the high reflectance foil 
for the white wall option of the IV vessel. The assumed reflectivities within the simulation are $0.4$ for stainless steel, $0.1$ for the 
black paint, $0.84$ for the white paint and $0.99$ for the high reflectance foil. \\
\\
These simulations result in maps (figure \ref{fig:MeanPEPlots}) showing in each 2 dimensional bin the 
number of detected PE/MeV for 1\,MeV electrons generated throughout this bin.  Each bin corresponds to the projection of the electron 
generation point on the IV surface. The blue color indicates regions where an energy deposition creates less detectable PEs, whereas energy 
deposited in red regions creates more PEs and the probability of detecting this energy deposition is higher at these points. The upper 
part of table \ref{table_PMTpos} summarises the results for the three different surface options (blank stainless steel, black painted surface and white painted surface with highly reflective VM2000 foil on the buffer vessel walls) for different PMT configurations with either $60$, $72$, $78$ or $108$ PMTs.\\

\begin{table}[]
  \centering
    \begin{tabular}{lcccc}\hline\hline
      Number of PMT     &60     &72     &78     &108    \\
      \multicolumn{5}{c}{\textbf{Mean PE/MeV}}\\
      stainless steel   &13.78  &16.66  &17.37  &18.88  \\
      black             &10.06  &11.96  &12.97  &13.22   \\
      white+VM2000      &30.42  &38.57  &38.77  &47.51  \\
      \multicolumn{5}{c}{\textbf{Fraction of events with 10 or less PE/MeV}}\\
      stainless steel & $52$\,\% & $33$\,\% & $37$\,\% & $25$\,\%\\
      black & $71$\,\% & $63$\,\% & $60$\,\% & $57$\,\% \\
      white+VM2000 & $2$\,\% & $2$\,\% & $1$\,\% & $6$\,\% \\\hline\hline      
    \end{tabular}
   \caption{Mean number of PE/MeV averaged over all 2-dimensional bins in the projection maps of 
local energy depositions for 1\,MeV electrons distributed uniformly within the IV volume. The influence of shadowing 
was estimated by determining the fraction of events which produce 10 or less PEs.}
  \label{table_PMTpos}
\end{table}

We considered painting the walls black in order to have the possibility of tracking muons with good spatial and time resolution. 
It was discarded as the simulation showed that more PMTs were needed compared to the blank steel surface. On average, the higher 
reflectivity of the white walls resulted in an increase of detected PE/MeV by a factor of $2-3$ compared to stainless steel surfaces. 
Furthermore, the improved reflectivity makes it possible to achieve a large amount of light collection ($\approx 39$\,PE/MeV) with only one ring of side PMTs on the veto vessel wall (compared to the configuration with $72$ and $78$ PMTs). Increasing the number of PMTs and changing their orientation (configuration with $108$ PMTs) shows a less homogeneous detector response (see figure \ref{fig:MeanPEPlots} for the IV bottom, where the inhomogeneity at the buffer feet is obvious).\\

The utilized option for the IV is the configuration with $78$ PMTs, white painted walls and VM2000 foil covering the buffer vessel. In 
this layout there are a total of $42$ PMTs on the bottom of the veto vessel, $12$ at the mid way point on the side wall and $24$ on the top of the veto vessel. This option was chosen because of the adequate optical coverage in combination with minimum shadowing effects (which occurs when one PMT covers another, reducing the field of view) and a high PE/MeV value. This value of $39$\,PE/MeV directly translates into a good ability for detecting corner clipping muons (track lengths around a few cm in the IV vessel), because they create at least $70$\,PE/cm. Also the utilized option has only $1$\% of the $1$\,MeV events producing less than $10$ photoelectrons (the comparison can be found in the lower part of table \ref{table_PMTpos}), while increasing the number of PMTs leads to a larger fraction of events below $10$ PE/MeV due to shadowing effects and is thus less efficient in detecting corner clipping muons.\\ 

\section{Determining rejection efficiencies}
\label{section_efficiency}

In the subsequent step MUSIC was used to propagate the sea-level muon distribution through a topographical map of 
the Chooz site in order to obtain the muon energy spectrum and angular distribution at the far detector underground site. 
The momenta of the muons underground were given as an input to a GEANT4 simulation in order to determine the rejection 
efficiency of the IV. Within the simulation a muon was placed at the detector center and the momentum was followed backwards 
to the simulations boundaries. At the boundary a $20$\,m$\times$33\,m rectangle was created perpendicular to the muon track and 
the muon was placed randomly on this rectangle and from this point it was propagated towards the detector with the original momentum. 
Muons generated on this rectangle can produce secondary particles which enter the detector and deposit energy therein. Enlarging the 
rectangle is not useful, because the secondary particles do not reach the detector volume any more.  In total around $10^7$ primary muons 
were simulated with this method. In $7$\,\% of the cases the detector was directly hit. Another $5$\,\% are events where the muon misses 
the detector but secondary particles deposit energy within. We used a MC data set that has $1.2\times 10^6$ events with an energy deposition 
in the detector (corresponding to approximately 8 hours of detector live time) to estimate efficiencies.\\

To quantify the IV's rejection abilities, two different definitions of rejection efficiency are used. First we define 
the muon rejection efficiency $R_{\mu}$ as

\[R_{\mu} = \frac{\rm{Number\,of\,recognised\,muons\,in\,the\,IV\,with\,energy\,deposition\,in\,the\,ID}}
{\rm{All\,muons\,with\,energy\,deposition\,in\,the\,ID}}.\]

\begin{figure}[]
  \begin{center}
       \includegraphics[width=10cm]{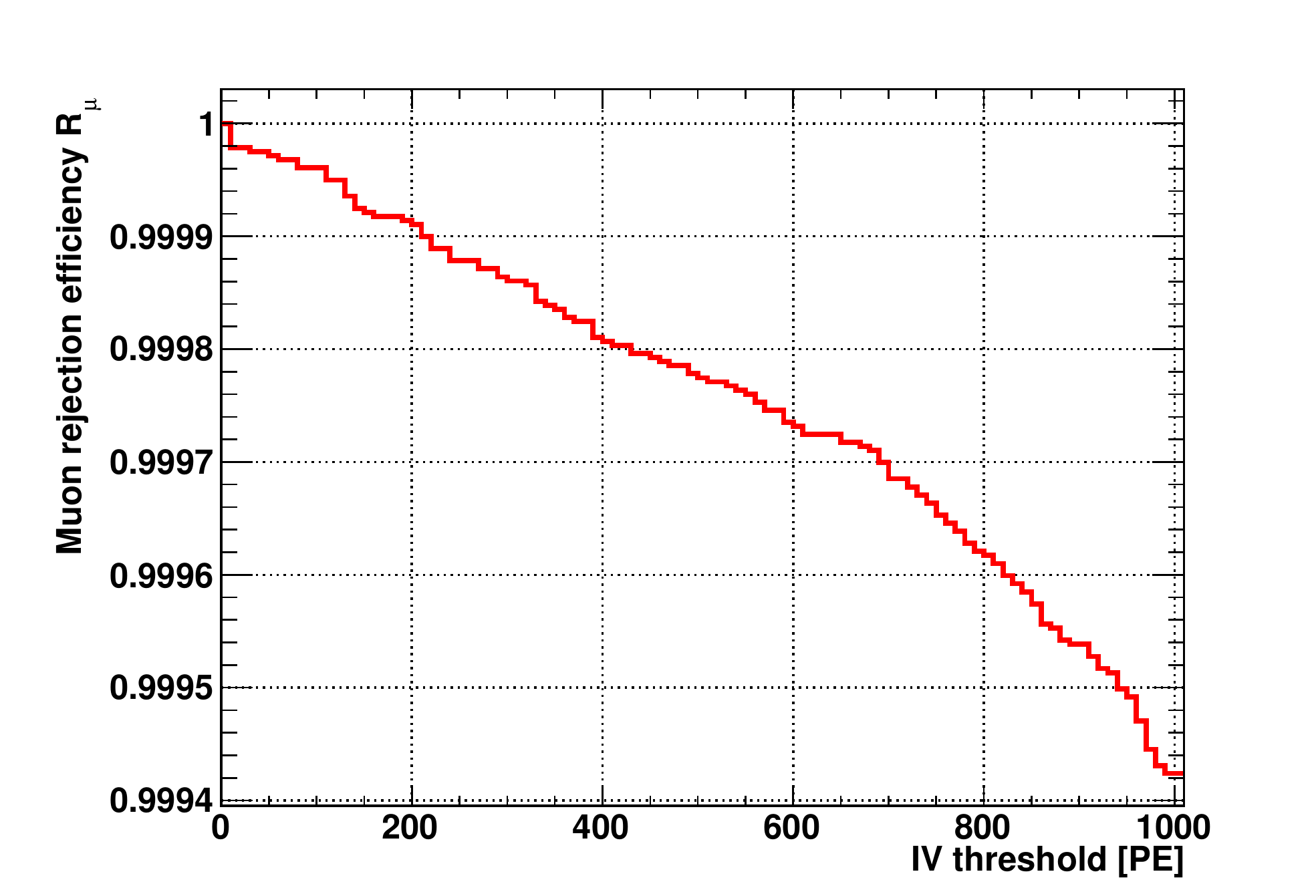}
    \caption{Muon rejection efficiency $R_{\mu}$ as a function of threshold in PE. Taking a threshold of $500$\,PEs the rejection efficiency is 
$99.978$\,\%.}
    \label{IVMuEfficiency}
  \end{center}
\end{figure}

This rejection efficiency reflects the IV's ability to tag and reject events that would contribute to the spectrum 
of all events (correlated as well as uncorrelated). A muon is considered as recognised if the number of PEs is above a given threshold. 
The effect of the threshold can be seen in figure \ref{IVMuEfficiency}. An efficiency of $99.978\pm 0.004$\,\% (at $90$\,\% C.L.) 
can be reached with a threshold of $500$\,PE. Going to lower thresholds is not feasible, because radioactivity from the 
surrounding rock and materials leads to high rates at lower thresholds.\\

A second definition is used in order to determine the efficiency for rejecting correlated, neutrino-like events, where a correlated event is 
defined as a visible energy deposition of $200-3500$\,PE (corresponding to $0.7-12$\,MeV) in the NT or GC that is followed by a visible energy 
deposition of $1700-3500$\,PE ($6-12$\,MeV) within $200\,\mu$s. We define the efficiency $R_\mu^c$ to reject muon-induced correlated events as

\[R_\mu^c = \frac{\rm{Number\,of\,correlated\,events\,recognised\,in\,the\,IV}}
{\rm{Number\,of\,all\,correlated\,events}}.\]

A correlated event is recognised if the time delay between it and the muon is less than $1000\,\mu$s. The specific value of the veto time was 
chosen to be long compared to the mean capture time on gadolinium ($30\,\mu$s) and on hydrogen ($180\,\mu$s). This veto time window introduces a dead time of $4$\% for the far detector and approximately $40$\% for the near detector. Thus this efficiency reflects the IV's ability to reject muon correlated events like spallation neutrons or cosmogenic isotopes with short life times compared to the veto time window.\\ 

In the simulation (using a threshold of $500$\,PE) all observed $224$ correlated events were recognised by the IV. Increasing the threshold to $1000$\,PE 
would lead to one unrecognised correlated event (having deposited $852$\,PE in the IV). Using a poissonian one-sided confidence interval, 
the efficiency for rejecting correlated events can be estimated to be $>98.98$\,\% (at $90$\,\%\,C.L.). This would lead to a maximum of $7$ 
unrecognised correlated events (without the IV this would be $645$) compared to $50$ expected inverse beta decay events in the detector per day. This result shows the importance of the IV for the feasibility of the experiment. 
The muon induced background rate of correlated events which would dominate the neutrino signal otherwise is now suppressed sufficiently with the inclusion of the IV.\\

The $7$ events per day is only the upper limit and the performance of the IV is expected to be even better. This was shown by a MC without tracking of scintillation light where much higher statistics for correlated background events can be reached \cite{DCProposal,SimGrieb}. In this simulation only one correlated background event in $43$\,h was found, corresponding to a rate of $0.55$ correlated events per day not being vetoed.

\section{Conclusion}
\label{label_conclusion}
In order that the Double Chooz experiment is able to measure the third mixing angle $\theta_{13}$ to sufficient sensitivity, a cosmic muon 
detector is needed to reject cosmic muons and any correlated event induced by them. For this purpose a proper cosmic muon 
background rejection scheme was reviewed using a veto detector. By changing the optical surfaces, positions and distributions of PMTs within 
the detector simulation, different designs and layouts could be compared. In the chosen design the veto vessel has white painted walls, the 
buffer vessel is coated with VM2000 highly reflective foil and the IV is instrumented with 78 PMTs. With this design a sufficient number of 
$39$\,PE/MeV was achieved. In addition, the simulations show that cosmic muons can be easily tagged (efficiency $99.978\pm 0.004$\,\%) 
and the correlated background due to muons can be rejected with an efficiency of $>98.98$\%. The inner muon veto for the far detector is already 
built and in 2011 a first indication for a non-zero value for $\theta_{13}$ was obtained, where the tagging of cosmic muons with this 
veto detector played a crucial role. The IV helps to reduce the correlated background due to spallation neutrons and $^9$Li down 
to $3.13$ events per day which is $3.7$\,\% of the measured neutrino signal \cite{DC1stPub}. While the current first phase of the Double 
Chooz experiment uses the far detector only, the near detector hall is currently being excavated. This detector will also 
be equipped with an IV and the detector installation is planned to happen in 2013.

\section{Acknowledgments}
\label{label_acknowledgments}

The authors want to thank the Double Chooz IV and Monte Carlo groups. This work was supported by the Deutsche Forschungsgemeinschaft through 
the Transregional Collaborative Research Center SFB/TR 27 ''Neutrinos and Beyond''. T.L. holds a ''Junior-Stiftungsprofessur der Carl-Zeiss-Stiftung''.


\end{document}